\documentclass{llncs}
\usepackage{graphicx}
\usepackage{color}
\usepackage{textcomp}
\usepackage{paralist}
\usepackage{url}
\usepackage{multirow}
\usepackage{xspace}
\usepackage[pdfborder={0 0 0}]{hyperref}
\usepackage{multicol}
\usepackage{caption}

\begin{document}

\title{Natural Notation for the Domestic Internet of Things}

 \author{Charith Perera\inst{1,2} \and Saeed Aghaee\inst{3} \and Alan Blackwell\inst{3}}

 \institute{Research School of Computer Science (RSCS) , The Australian National University, Canberra, ACT, Australia\\
 \and Faculty of Maths, Computing and Technology, The Open University, UK\\
\email{charith.perera@ieee.org}
 \and Computer Laboratory, University of Cambridge, Cambridge, United Kingdom\\
  \email{\{saeed.aghaee, alan.blackwell\}@cl.cam.ac.uk}
 }

\maketitle
\begin{abstract}
This study explores the use of natural language to give instructions that might be interpreted by Internet of Things (IoT) devices in a domestic `smart home' environment.
We start from the proposition that reminders can be considered as a type of end-user programming, in which the executed actions might be performed either by an automated agent or by the author of the reminder.
We conducted an experiment in which people wrote sticky notes specifying future actions in their home. In different conditions, these notes were addressed to themselves, to others, or to a computer agent.
We analyse the linguistic features and strategies that are used to achieve these tasks, including the use of graphical resources as an informal visual language.
The findings provide a basis for design guidance related to end-user development for the Internet of Things.

\keywords{Internet of Things, smart home, end-user programming, user study}

\end {abstract}

\section {Introduction}
\label{sec:intro}
In this paper, we investigate how people might program the `smart home' domestic technologies enabled by the Internet of Things (IoT).
However, rather than start by creating another new end-user programming (EUP) language, we study how people give instructions in the home as an existing natural task.
Using the familiar sticky note as an experimental device, we conducted an experiment in which we asked people to write sticky notes requesting that things should be done in their homes.
We wanted to compare routine requests to other people, in comparison to communication with a programmable smart home. As a control condition, we compared both of these to the use of sticky notes as a reminder to oneself.

\section {Background}
\label{sec:background}

A major category of applications for IoT devices are `smart home' scenarios, in which domestic tasks are automated by defining policies or scripts.
These scenarios appear attractive to technical enthusiasts, but actual home automation currently faces a key obstacle, in the ability to control communication between devices. Although there are many categories of automated domestic appliance (e.g. floor-cleaning robots, bread-makers, video recorders), their automated behaviours are carried out by a single device.
In such cases, the functional capabilities of the device are determined by the manufacturer, with the user only needing to customize it to suit the configuration of their own house or daily schedule.

In contrast, the most challenging instances of EUP for IoT devices are those that involve information exchange -- between devices and users (e.g. reminders), between devices and services (e.g. automated orders for home supplies), or between devices themselves. Many domestic information exchange tasks are activities that could in principle be carried out directly by the user, reading information from one device or context and applying it to another.
For example, when you notice that the laundry powder box is nearly empty, you remember (perhaps) to buy powder next time you are shopping. However, all these little tasks consume the limited resource of human attention.
One way of defining domestic EUP is that it allows people to optimize their allocation of attention, choosing between immediate direct manipulation and reminders or automation of future behaviour~\cite{Blackwell2009,Dey2006}.

There are numerous popular examples of end-user programming services for the Web that demonstrate this strategy.
For example, IFTTT\footnote{\url{http://ifttt.com}} allows users to specify future behaviour as an if-this-then-that policy. 
A typical IFTTT `recipe' might be triggered (if) every time a photo is taken on your iPhone (this), and then automatically upload it to your Twitter feed (that).
IFTTT is a simple and practical EUP system~\cite{Ur2014} that automates future actions. 
In this paper, we explore a familiar category of home behaviour in order to gain insight to the opportunities for creating IFTTT-style automation services for the home.
We focus on tasks that involve attention investment -- where paying attention to something in advance, in order to define a policy, will save attention in future.
The everyday term for this kind of information exchange is a reminder -- creating a mechanism that transfers information at a time in the future.
Similar rule-based systems include Tasker\footnote{\url{http://tasker.dinglisch.net}}, Atooma\footnote{\url{https://play.google.com/store/apps/details?id=com.atooma}}, and Locale\footnote{\url{http://www.twofortyfouram.com}}.

In terms of EUP research, there is a well-established strategy for using everyday descriptions of an automated task as a way of gaining insight into programming system design.
In the method employed in Myers' Natural Programming project~\cite{Myers2004}, typical studies recruit samples of representative users who are asked to describe specific types of program behaviour in their own words.
The key to design of valid natural programming studies is to ensure that the experimental tasks correspond to the intended application domain, and that the programming `environment' in which the natural language description is collected properly represents the cognitive demands of the programming situation.
For example, natural programming studies do not usually proceed by asking participants to give a verbal description to the experimenter, because human conversation relies on substantial elements of common ground and interpretive ambiguity that are not available in programming languages.

Our goal in this study was therefore to define an experimental paradigm that could be used to study investment of attention in the definition of domestic information exchange policies, in a manner that offered external validity with regard to the context of everyday home management.
We chose to focus on the sticky note (a generic term for the product category introduced by Post-It\texttrademark). One of the major uses of the sticky note in domestic contexts is indeed to implement reminders. People often write sticky notes as reminders to themselves, and place notes in a context where they expect future information exchange to be valuable -- on the front door, on their keys, on their wallet and so on. 
In shared houses, people also write sticky notes and place them in contexts where they wish to remind other people of particular policies or requests for action. Our study generalizes beyond these familiar cases, to use the sticky note as an experimental proxy for a domestic EUP language, where the reader of the note is not yourself or another resident of your house, but an automation service that will take responsibility for future information exchange.

\section{Related Work}
\label{sec:related}

There is a long tradition of exploring the potential design space of `smart home' technologies through study of natural behaviours in the home (e.g.~\cite{Tolmie2002}), and investigating the ways in which families respond to the opportunity to program and configure existing digital technologies~\cite{Rode2004}.
Sticky notes are often included in the kit of materials for cultural probe studies (e.g. Graham et al.~\cite{Graham2007}), and the sticky note metaphor has been literally rendered in home technology probes (e.g. Hutchinson et al.~\cite{Hutchinson2003}).
The refrigerator door, as a location where sticky notes and other papers are placed, is a regular focus of smart home technology, both as a versatile augmented display surface for research attention~\cite{Taylor2006} and more literally in commercial products such as the Samsung WiFi-enabled fridge\footnote{\url{http://www.samsung.com/us/topic/apps-on-your-fridge}}.
One can imagine that sticky notes themselves might be used as an EUP technology in future (e.g. Tarkan et al.~\cite{Tarkan2010}), and there are many previous systems that have augmented sticky notes in ways that might achieve this (e.g. as reviewed by Mistry et al.~\cite{Mistry2009}).

Our goal in this research is to understand better the ways that people express themselves when making requests or reminders in the smart home context.
This can help to design more natural EUP systems, as in the work of Myers et al. It can also be applied as a basis for Natural Language Interfaces (NLI). Many companies are currently deploying NLI for simple query and status applications, especially in mobile apps such as Google Now\footnote{\url{http://www.google.com/landing/now/}} and Siri\footnote{\url{https://www.apple.com/ios/siri/}}.
For instance, Google Now has recently published an advertisement\footnote{\url{https://twitter.com/googleuk/status/525194969238478848}} that tackles a similar use case for smart homes.
The Microsoft Cortana\footnote{\url{http://www.windowsphone.com/en-us/features-8-1}} NLI has recently been exploited in INSTEAON\footnote{\url{http://www.insteon.com/}}, a mobile app that aims at creating a natural language interface for IoT.

Nevertheless, when considered as a programming interface, NLIs have numerous disadvantages.
The ambiguity and lack of precision in natural language remains a challenge~\cite{Petrick1976}, with more detailed instructions often less efficient than a concise formal notation~\cite{Dijkstra1979}.
Analysis according to Cognitive Dimensions of Notations~\cite{Blackwell2003} identifies that speech-based interaction
\begin{inparaenum}[(1)]
\item poses constraints on the order of doing things (premature commitment),
\item conceals information in encapsulations (poor visibility),
\item doesn't allow changes to made decisions (high viscosity), and
\item obscures links between entities (hidden dependencies).
\end{inparaenum}
The sticky note offers an opportunity to study natural language interaction in a written context which is routinely augmented with visual cues.

\section{Structure of the Study}
\label{sec:structure}

We designed six different use case scenarios for presentation to participants in the study (see Table~\ref{tab:scenarios}).
Each scenario depicts a familiar problem likely to be faced in the home, requiring future action either by yourself (recording a reminder), by someone you live with (delivering a request) or potentially by an intelligent agent of some kind (defining a program or script). Participants were asked to write a sticky note to `solve' the problem mentioned in each scenario, in a manner that would implicitly result in the generation of a speech act such as reminder, request, or program.

\begin{figure}[h]
    \begin{center}
    \includegraphics[trim = 1mm 10mm 22mm 15mm, clip,width=1.0\textwidth]{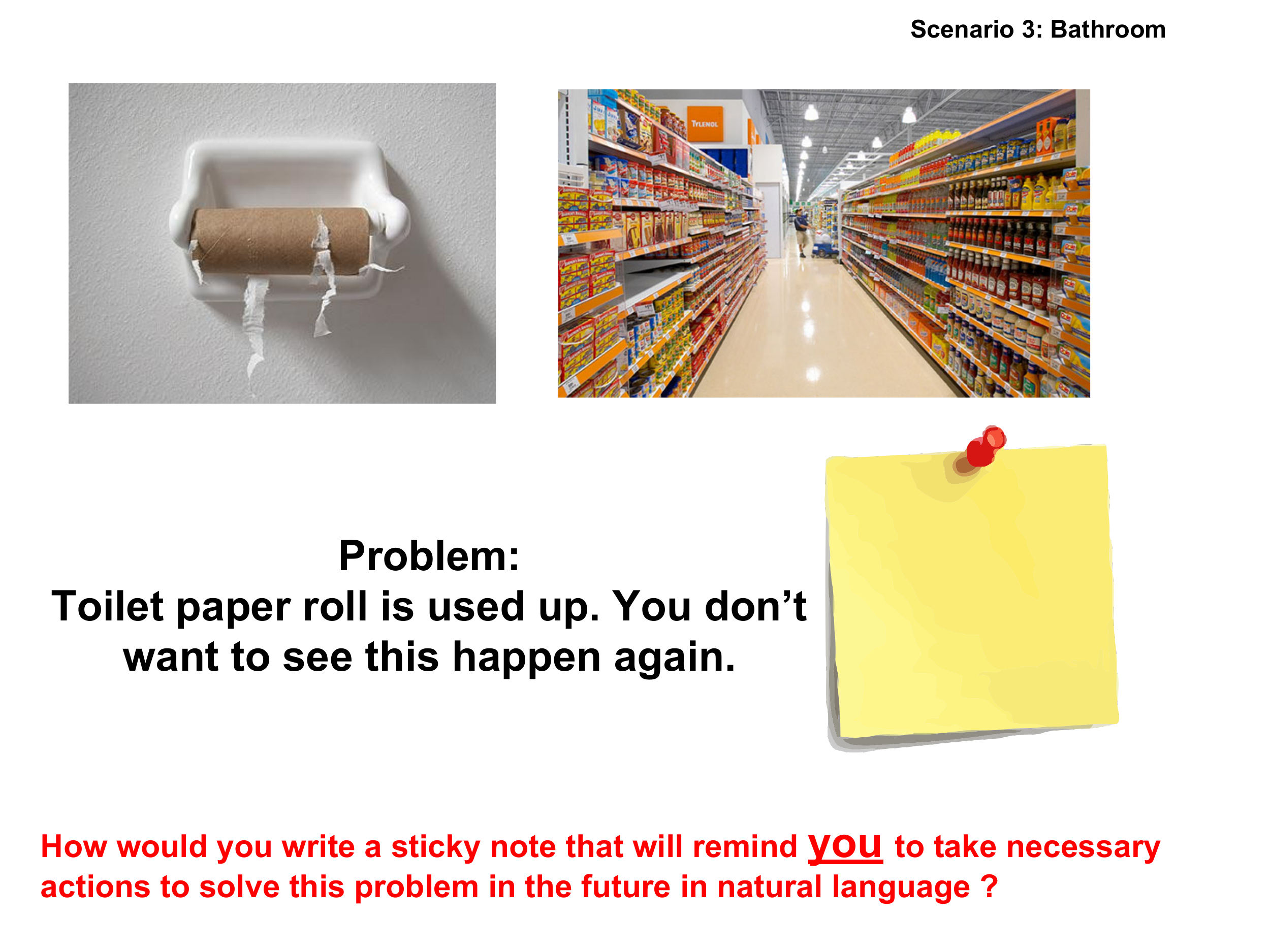}
    \caption{Sample use case scenario sheet}    
   \label{fig:case}
    \end{center}
    \vspace{-20pt}
\end{figure}

We presented each scenario on an A4 sheet, using a minimum number of words in order to encourage participants to choose their own phrasing for the resulting speech act.
Instead of verbal description, the problem context was illustrated as far as possible using images from which the nature of the problem could be inferred.
An example is shown in Figure~\ref{fig:case}.
Our goal was to provide sufficient information to draw attention to the nature of the problem, without including any direct phrasing that might be incorporated in the participant responses.
As with other studies in natural programming, we wanted participants to think about how they would approach the problem, using their own words and presentation mechanisms to write the sticky notes without being influenced by our descriptions and instructions. 

Each A4 sheet included a printed rendering of a sticky note.
We pasted a real sticky note on top of this printed sticky note before handing the survey sheets to each participant.
Use of a real sticky note increased the naturalness of the experiment, allowing participants to imagine that these notes might be placed in their own home (several participants made unsolicited comments indicating that they did indeed imagine particular locations in which the note might be placed).
Attaching a sticky note to the instruction sheet was also convenient for data collation and analysis, in that we could remove the sticky notes and arrange all six on a single page for each participant.

The six scenarios covered a range of home activities, designed to take place in different contexts within a typical house, covering different frequencies and durations of activity, and representing different degrees of complexity and cost. These are summarized in Table~\ref{tab:scenarios}.

\begin{table}[h]

\caption {The problems described in each scenario} 
\vspace{-10pt}
\label{tab:scenarios}

\begin{center}
\begin{tabular}{|l|p{8cm}|}
\hline
\textbf{Context} & \textbf{Problem}                                                                                                                                                    \\ \hline
Laundry          & Washing Machine filter is clogged. This happens roughly every 3 months                                                                                              \\ \hline
Kitchen          & You have prepared food for your kids and about to leave your house. You won't come back until late. Leftover food can be spoiled if it is not placed in the fridge. \\ \hline
Bathroom         & Toilet paper roll is used up. You don't want to see this happen again.                                                                                              \\ \hline
Garage           & It is summer!!!.. Your parents have asked to bring your weed eater when you visit them next time. Every summer they need your weed eater to cut their lawn.         \\ \hline
Living Room      & Some relatives come to visit every few months... Your house is usually a mess                                                                                       \\ \hline
Garbage Bins     & You always forget to put garbage bags into outside bins located in front of your house so the council will pick them up on Mondays                                  \\ \hline
\end{tabular}
\end{center}
\vspace{-24pt}
\end{table}

In each scenario, we asked the participants to write a sticky note to solve the illustrated problem. We created three variants as follows, each addressed to a different person who will carry them out (the `addressee'): 

\vspace{4pt}
\emph{How would you write a sticky note that will}
\vspace{-6pt}
\begin{itemize}
\item (version 1) remind you
\item (version 2) remind someone you are living with
\item (version 3) be interpreted by a machine (an intelligent robot or something that can read sticky notes) 
\end{itemize}
\vspace{-10pt}
\emph{to take necessary actions to solve this problem in the future in natural language?}

We created six sets of experimental materials, each using the same scenarios, but with different combinations of instructions, balanced across participants.
Our main within-subjects research question relates to the effect of addressee, so the three different addressee conditions were placed to minimize order effects.
Issues with participant recruitment and withdrawal led to slight variations (15.9\% in sets 1, 5 and 6, 14.3\% in set 2, 20.6\% in set 3, and 17.5\% in set 4). 

In addition to completing sticky notes for the six scenarios, participants completed a short questionnaire: gender, age group, profession, education, number of hours spend using computational devices, and previous programming experience. A final debriefing question asked for any further comments on the experiment. 

\vspace{-8pt}
\section{Summary of Data Collected}
\label{sec:summary}
\vspace{-8pt}

\begin{figure}[h]
    \begin{center}
    \includegraphics[scale=1.1]{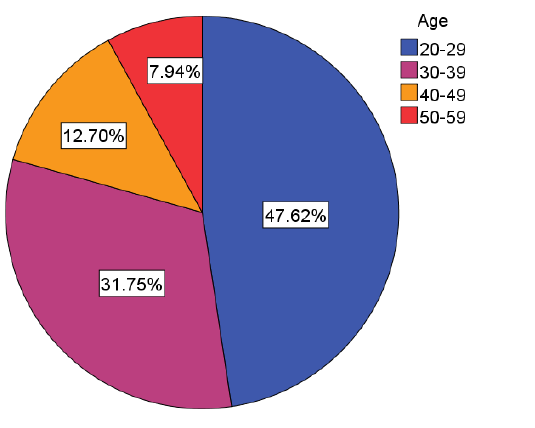}
    \vspace{-15pt}
    \caption{Age distribution of participants}    
   \label{fig:participants}
    \end{center}
    \vspace{-20pt}
\end{figure}

We recruited 63 participants in Canberra, Australia. All had good working knowledge of English, either native or fluent speakers. No direct compensation was provided for participating in the study. 38 were male (60\%) and 25 female (40\%).
Figure~\ref{fig:participants} shows the age distribution. All participants were met either in their office or at home. Most completed the tasks and demographic questionnaire in 5-10 minutes, although a small number (n=4) took 15-20 minutes.

In response to the final debriefing question, two participants requested further information about our research.
Two noted that they did not know how the machines would work.
One noted that she was not a native English speaker.
One questioned the practicality of using sticky notes for these scenarios.
Several participants also used the sticky notes themselves to provide additional explanation to the researchers.
This included information on where they might place the sticky note, for example adding \emph{In front of the door} to a sticky note \emph{``Place the food in the fridge''} (in response to the kitchen scenario).
A few participants wrote an alternative method that they would prefer to writing sticky notes, for example \emph{I would probably put an alert in my calendar} (in the garage scenario).

Based on the questionnaire responses, we constructed a measure of prior technical experience.
Where a participant stated that they were familiar with multiple programming languages, they were considered to be technically experienced.
In cases where the participant reported some experience of a programming language used in school, they were only placed in the technically experienced category if they were in a technical profession, or had a degree in a technical field.
Although this is a relatively coarse criterion, it was less intrusive and faster to administer than more elaborate programming aptitude tests.
It is possible that some participants were mis-classified as a result, but since the classification was done without reference to responses, we treat this as an unbiased source of experimental variation. Demographic data showed that the technical experience construct was correlated with gender (8 females and 26 males).

\section{Analysis Method}
\label{sec:method}

\subsection{Linguistic Analysis}
\label{sec:linguisticmethod}

Linguistic analysis involves studying the grammatical structures and meaning of a language sample.
In the context of the smart home, our goal was to develop an understanding of how the communication of information between people and smart devices is linguistically framed.
Linguistic framing offers insights into the mental model of users as well as their communicative practices. These, in turn, provide valuable guidelines for designing ``natural'' programming languages that conform to those mental models and practices~\cite{Myers2004}.

Our analysis corpus consisted of 374 sentences extracted from the sticky notes. It was not possible to use automated tools due to the ambiguity caused by frequently missing punctuation in the notes. Note that although this corpus is relatively small compared to those used in automated text analysis, our goal was to make a descriptive analysis rather than derive statistical training data of the kind that would be necessary in creating a full natural language interpreter.

We proceeded with the analysis at three levels of Syntactic, Semantic, and Pragmatic, which correspond to the analysis of, respectively, form, meaning, and context of the language used by the users to frame the reminders.

At the syntactic level, we first assessed the frequency of noun phrases compared to sentences in the corpus.  Furthermore, we manually classified all the sentences according to their grammatical structure. Our classification of sentences involves the grammatical differences between Declarative, Interrogative, and Imperative sentences. Declarative sentences state a fact and syntactically consist of a subject that normally precedes a verb. Interrogative sentences are constructed in the form of yes-no questions or wh-questions (i.e., questions framed using an interrogative word such as ``who'', ``which'', ``where'', and ``how'').
Imperative sentences give a command or make a request and have an understood but not always stated ``you'' as the subject.

We applied the standard semantic analysis method of n-gram modelling to estimate the probability of a given sequence of words. An n-gram is simply a sequence of n words that appear in the corpus in the same order immediately one after another.
Using the SRI Language Modelling Toolkit~\cite{Stolcke2002}, we built a list of most recurrent uni-grams (1-grams) and bigrams (2-grams). 

We refined the n-gram list by removing common n-grams that had no direct reference to the task (e.g. ``in the'') as well as the phrases that were directly prompted by the scenario description presented in the experiment (e.g. ``washing machine'', ``weed eater'' and ``fridge'').
We then grouped the n-grams according to semantic similarity, using the ConceptNet word association API~\cite{Speer2012}.
ConceptNet uses a rich commonsense knowledge base to provide an accurate measure of similarity between two concepts.

Extracting context-dependent meaning from a language sample is, however, beyond semantic analysis and is dealt with at the pragmatic level. Speech acts are used to express a certain attitude and are a central point of pragmatics. Hence, we considered the five classes of speech acts proposed by Searle~\cite{Searle1976}, and classified the sentences into one of five speech act classes as follows:

\begin{description}
\item [Representatives:] The speaker asserts his or her belief of something that can be evaluated to be true or false (e.g., a doctor's diagnosis about the presence of a disease).

\item [Directives:] The speaker expects the listener to take a certain action as a response such as by asking a question or making a request.

\item [Commissives:] The speaker commits to a future action, for instance, by making promises or threats.

\item [Expressives:] The speaker expresses his or her attitude and psychological state (e.g., thanking or apologizing) to the listener.

\item [Declaratives:] The speaker changes the status or condition of the reality, for example, by pronouncing a marriage.
\end{description}

\subsection{Visual Language Coding}
\label{sec:visualmethod}

The visual/typographic elements of participants' responses were coded using the graphical language approach introduced by Bertin~\cite{Bertin1983}, extended by Engelhardt~\cite{Engelhardt2002}, and applied to user interfaces by Blackwell~\cite{Blackwell2013}.
This analyses the `graphic resources' that have been employed in a hierarchy of marks, symbols, regions and surfaces.
Each of these supports a range of semantic correspondences that allow different types of design application. In our experiment, we can consider the participants writing sticky notes as implicit designers, inventing their own graphical languages in response to the task.
The visual language analysis started with overall ink distribution, including multiple regions separated by whitespace (if any) and bounded regions formed by visual gestalt properties of alignment or regular containment.

The great majority of the marks made by participants were English alphabet letter forms, in most cases either uniformly uppercase, or uniformly lowercase (with appropriate grammatical capitalisation).
Some distinctive capitalisation was observed -- this is discussed below. A relatively small number of participants made no further use of visual language devices -- they wrote conventionally from left to right, starting at the top of the sticky note, and starting a new line when necessary.

Punctuation marks were coded separately, as were other conventional symbolic forms such as emoticons and logos.
A small number of respondents included pictorial elements, although these included meta-communication with respect to the task frame, such as a drawing of the door on which the sticky note should be placed. Other functional graphical devices included connecting lines, dividing lines between regions, and a small number of conventional typographic or diagrammatic forms such as tables and flowcharts with arrows and decision diamonds.

Visual language coding was performed by annotating colour scanned copies of all participant responses, marking every occurrence of the features discussed. This was done in two passes -- first open coding, in which all types of visual feature in the sample were identified, and then exhaustive coding of each occurrence of that type. Some qualitatively distinctive but infrequent features were noted for discussion.
Frequently occurring features were tabulated and transcribed into our statistical data set for each case in which they had been observed.

\section{Findings}
\label{sec:findings}

\subsection{Linguistic Structure}
\label{sec:linguisticfincdings}

In this section we make a linguistic analysis of the sticky notes. We consider features at several syntactic, semantic, and pragmatic levels: number of words (syntactic), sentence types (syntactic), request types (semantic and pragmatic), causality (semantic and pragmatic), and speech acts (pragmatic).

\subsubsection{Number of Words}

The sticky notes we used in the experiments were small in size, thus restricting the reminder messages to a limited number of words.
On average, the number of words per reminder message was 9.
Looking for differences between the participants, we found that the addressee of the sticky note has a statistically significant impact on the number of words, whereas technical experience (and gender) has no significant effect. The average number of words in a note addressed to yourself was 7.83, which was significantly less than those addressed to someone else (10.86) or to an intelligent machine (10.33) (ANOVA $F=(6.47,2), p<0.01$).

\subsubsection{Sentence Types}

Out of 374 sentences, we classified 271, 94, and 9 as imperative, declarative, and interrogative, respectively. Gender and technical experience do not impact the sentence type. However, we found a significant difference in the effect of addressee. Sticky notes addressed to an intelligent machine and the participants themselves have, respectively, the least (81) and the most number (100) of imperative sentences (Chi-Square test $p<0.05$).

\subsubsection{Request Types}
Table~\ref{tab:ngram} lists the most common n-grams. Based on the ConceptNet API, we put `remind', `remember', and `don't forget' in one category (similarity score$>0.9$), of `reminder' requests. While technical experience and addressee had no effect, gender was a significant factor: significantly more male participants (228 cases out of 374) used one of these (formal) ``request'' phrases (Chi-Square test $p<0.05$). 

We noted that the phrases `remember' and `don't forget' exhibit a double negation. We found that significantly fewer cases incorporate one of these phrases when the addressee is an intelligent machine, as compared to a human (someone else or the participants themselves). 

\vspace{-10pt}

\begin{center}
\captionof{table}{The most important n-grams (unigrams and bigrams)}
\label{tab:ngram}
\begin{minipage}[t]{.40\textwidth}

\begin{tabular}{|c|c|c|}

\hline
\textbf{N-gram} & \textbf{N} & Occurrences \\ \hline
put             & 1          & 69          \\ \hline
clean           & 1          & 63          \\ \hline
please          & 1          & 52          \\ \hline
if              & 1          & 50          \\ \hline
house           & 1          & 26          \\ \hline
when            & 1          & 24          \\ \hline
\end{tabular}
\end{minipage}
\begin{minipage}[t]{.40\textwidth}
\begin{tabular}{|c|c|c|}
\hline
\textbf{N-gram} & \textbf{N} & Occurrences \\ \hline
remember        & 1          & 20          \\ \hline
remind          & 1          & 19          \\ \hline
remind me       & 2          & 14          \\ \hline
remember to     & 2          & 13          \\ \hline
dont't forget   & 2          & 12          \\ \hline
\end{tabular}
\end{minipage}
\end{center}

%
%


\vspace{-20pt}

\subsubsection{Causality}

Another set of related recurrent n-grams are `if' and `when'.
These words have been used to indicate and express a cause or condition in a sentence. The results show that technical experience and having a machine as an addressee positively affects the use of these causal words in a sticky note (respectively 202/374 and 125/374 cases). Fewer of these cases were self-addressed (123/374 cases). There is a subtle grammatical difference between `if' and `when': `if' is used when the outcome is not certain, while `when' is used when it is certain. Overall, the frequency of `if' is more than double that of `when'.
We noted that in scenarios 1, 2, and 6, where the outcome depends on a condition that may hold, the number of `when' cases is considerably lower than that of `if' cases. In other scenarios (3, 4, and 5) the two phrases are almost equally utilized.

\subsubsection{Speech Acts}

As anticipated by the nature of the scenarios, the dominant speech act was of directive class ($296/374$ cases). Moreover, the addressee had a significant impact in this matter. The sentences are more likely (108/374 cases) to convey a directive speech act when addressed to an intelligent machine (Chi-Square test $p<0.05$). We found no commissive (promise or threat) and only one instance of expressive (expressing emotions) speech acts with a machine specified as the addressee. Also, we found no expressive speech act that was self-addressed.

\subsection{Graphical Resources}
The following discussion analyses the visual features observed according to their frequency across all cases, where each case represents a single sticky note. 255 out of 378 cases (67\%) included one or more of the visual features discussed above. Only 2 participants used no visual features in any note, writing in `plain' text. 

\subsubsection{Visual Regions}
45 cases divided the note into two distinct regions, always separated vertically. In 14 of these, a distinct region at the top functioned as a title, announcing the purpose of the note. In 19 cases, regions within the `body' area were separated either by a horizontal line, or an area of blank space, with a few distinguishing one region from another by different handwriting styles in each. A further 12 cases used the conventions of written correspondence, with a salutation and farewell at the top and bottom of the note. Two participants used more elaborate decoration -- highlighter pen overlay and `bang' circle.

Text arrangement within a region was most often conventionally left- and top-justified -- a natural arrangement for handwriting as it requires little advance planning. There were 11 cases in which the text block was aligned with a slant: of the left edge, of the text base lines, or both. A further 11 cases showed consistent centre-alignment of the individual text lines. This is surprising, given that it is difficult to achieve (the length of each line must be known before starting it), and does not have a clear semantic purpose.

Vertical left-alignment was used to indicate items in a list -- in 5 cases prefaced by a dash, in 4 with sequential numbers, in 3 with round bullets, and in 2 cases purely implicit with no marker. Nested left indentation, as in programming language source code, was used in 5 cases.

\subsubsection{Symbols}

As noted, alphabetic letters were usually lower case with conventional capitalisation. There were also 9 cases of idiosyncratic capitalisation, usually of selected nouns. Rather than title case convention, these seemed to apply to words marking key semantic concepts in the message (e.g. ``Date''), so may be related to computer idioms. Individual words were sometimes capitalised for emphasis (9 cases), or had an underscore added (7 cases)

After alphabetic letters, the most common symbol was use of an exclamation mark (26 cases). Parentheses were used to separate supplementary information in 2 cases, and long dashes to indicate pauses in 3 cases, a link in one case, and item markers as already mentioned. Algebraic symbols were used in 8 cases, often in ways that reflected typical programming language practice.

Smiley emoticons were used 6 times, in the conventional form with a circle around two eyes and mouth. There were 4 cases of other icons, using visual analogy to traffic signs, from a single participant. Finally, there were 7 cases in which a terminator symbol was added to close the message -- either an underline or flourish.

\subsubsection{Visual Semantics}
Overall, it is apparent that responses included many visual features beyond those that can be captured in a text transcription, or that correspond to semantic content of speech interaction. We saw no sign of `private' visual language. As a result, many of these could conceivably be used as design resources in multi-modal interaction systems, employing visual language devices that form a common vocabulary of written speech acts. 

However, it is possible that some of these features form a specialised visual vocabulary that would not be appropriate if imposed on end-users (or alternatively, a precise notation that might have to be taught to end-users in order for them to use a visual command language competently). We therefore carried out a statistical analysis of the distribution of all these features, with particular attention to whether the `speaker' (the participant who wrote the sticky note) or the `listener' (the addressee of the note) suggested that a technically specialised visual vocabulary was being used.
\subsubsection{Visual Pragmatics}

Do people adjust the visual language grammar they use when they are writing a sticky note addressed to an intelligent machine, rather than another person? As with other EUP research, we might expect that people with programming experience are more likely to have experience of specific conventions of language use derived from programming.

We have three hypotheses: 
\begin{description}
\item [H1)] that there is an identifiable subset (or `dialect', perhaps) of visual language features that are more often used when addressing machines rather than people 
\item [H2)] that there is a complementary set of visual language features that are more often used when addressing people rather than machines
\item [H3)] that people with prior experience of programming are more likely to use an identifiable subset of visual language features when addressing machines
\end{description}

(null hypotheses are that there is no difference in frequency of features resulting from addressee of the sticky note or technical experience of the writer)

During coding, we observed several visual features that appeared characteristic of program source code layout rather than other handwriting or print conventions: use of nested indentation, use of ordered lists of steps, and use of algebraic symbols. We counted all cases in which any of these three conventions had been used, finding 52 cases. Of these, 32 were addressed to an intelligent machine, with only 10 addressed to someone else and 10 self-addressed (Chi-Square $p<0.001$). The people writing these were more likely to have technical experience -- 43 out of 52 cases (Chi-Square $p<0.001$). Nested indent was particularly likely to be used by those with technical experience (26/27 cases, $p<0.001$), and to be addressed to a machine (20/27 cases, $p<0.001$). Ordered lists were also more likely to be addressed to a machine (11/16 cases, $p <0.05$), although sample size is too small to state that there is a difference based on technical experience. Algebraic symbols were more likely to be used by those with technical experience (20/26 case, $p<0.05$), but we cannot say whether these were more likely to be addressed to a machine. Algebra was unlikely to be used in the living room scenario (1/26 cases), which involved no numeric values. Lists were most frequently used in the living room and laundry scenarios (12/16 cases).

We observed several visual features that appeared characteristic of informal correspondence: use of salutations, emoticons, exclamation marks and visual emphasis. We found 102 cases, of which only 19 were addressed to a machine (Chi-Square test $p<0.001$). People with technical background were less likely to use these human-like visual conventions (46/102, $p<0.05$), in particular when addressing machines, although we did observe some cases (7 of 19, n.s.). The largest number of these features appeared in notes written to someone else (50/102 cases, $p<0.001$), especially the use of salutations (18/20, $p<0.001$). Emoticons were not used at all when addressing machines, and most likely to be used when addressing someone else (6/8, $p<0.05$). Exclamation marks were relatively unlikely to be used when addressing machines (12/65, $p<0.05$), although these cases were evenly split between technical and non-technical writers. Writers with technical experience were relatively unlikely to use visual emphasis (6/33, $p<0.001$) with only one of these cases addressed to a machine.

We also observed trends in the overall visual structure of the notes created by people with technical and non-technical backgrounds. Those with technical background were less likely overall to divide the note into separate regions (23 out of 72 cases, Chi-Square test $p<0.001$) and less likely overall to place a context or mode title at the top of the note (15 out of 45 cases, $p<0.01$). These may be general habits derived from use of sticky notes in technical work, because we saw no evidence that these overall visual structures were more or less common according to either the addressee, or the task scenario.

Overall, we find that there are differences in visual language features that are used by people with technical and non-technical backgrounds when writing sticky notes (rejecting the null hypothesis for H3), and that there exist particular sets of visual language features likely to be used when addressing machines (H1) and people (H2).

\section{Implications for Design}
\label{sec:implications}

While general natural-language programming remains a challenging future ambition for EUD, our focus on the specific application domain of IoT and on the sticky note as a constrained multimodal information device suggests new opportunities for EUD.

We have shown that the graphical resources of the sticky note complement natural language understanding, by allowing the use of visual language cues that establish the context for instruction, drawing on a number of commonplace graphical conventions.
This represents an opportunity for simple multimodal interfaces to support EUP in this domain, which has remained untapped.
Despite the fact that systems such as IFTTT, Atooma, and Locale, incorporate a visual iconic language (e.g., rules and actions are visualized as icons), they do not support a multimodal interface capable of receiving both natural and visual language inputs in a complementary manner. 

We have also identified a number of ways in which people make allowance for interpretation when addressing others (either machine or person) rather than themselves. This helps us to understand the cognitive effort involved in `reminder' tasks, which inherently involve cognitive effort to anticipate future state. For example, the distinction between `remember' and `don't forget' (the case of double negation) requires implicit theory of mind judgements~\cite{Jespersen1992}.
Our results show that, while such anthropomorphic considerations are always implicit in reminders, and sometimes involve expressive speech acts with emotional elements, communications intended for interpretation by a machine are far less likely to include such elements.
The majority of existing smart home systems are used for automation solutions and thus naturally mediate communication between humans and machines at the user interface.
Nonetheless, some interesting use cases have emerged beyond machine automation and have been implemented to enhance human-to-human communication such as in Remind'em app\footnote{\note{http://www.remindem.in/}}.
Hence, this subtle difference in the attitude of end-users towards the communication target should be taken into account while designing the user experience for smart home systems.

As might be expected, people with technical experience bring this to bear in speech acts directed toward a machine. 
This results in more detailed specification, with syntactic and semantic forms that resemble programming language constructs. 
It is interesting to note that this resemblance covers both textual and graphical modes -- a consideration that should be taken into account when designing `natural' interfaces.
On one hand, it is a safe strategy to design an EUP system in compliance with the non-programmer mentality~\cite{Aghaee2012} (e.g., high simplicity and low expressive power), and on the other, offering a high level of expressive power can also be beneficial for non-programmers as well.
Given a high level of expressive power, programmers will be able to write high quality scripts that can be reused by many end-users.
Moreover, a study by Lucci et al ~\cite{Lucci2014} shows that expressive power is also a factor for non-programmers in deciding which smart home automation service to use.

Finally, our combined visual/textual analysis demonstrates the devices used to accommodate semantic modes such as `if' and `when', as a component of the conventional reminding and instruction speech acts accomplished with sticky notes. While a trigger such as `relatives paying a visit' is accommodated in emerging event-based mashup paradigms (e.g. IFTTT), the attention investment required for modal reasoning about temporal contexts such as `when' may involve more sophisticated combinations of natural language and other notational devices.
Existing rule-based systems (e.g., IFTTT, Tasker, Atooma, and Locale) seem to disregard these subtle differences between these semantic modes and their effects on the design of the overall user experience.

\section {Conclusion}
\label{sec:conclucion}
In this paper, we have reported an experiment that explored the ways in which people express themselves, when specifying domestic tasks of the kind that will require exchange of information between IoT appliances and services. We used the familiar sticky-note, which has often been a focus of research for understanding everyday information technologies. The sticky note offers good external validity for experimentation in this area, as a mundane information technology that supports the same kinds of reminder function identified in the attention investment model of EUP. We have carried out a rigorous linguistic analysis of this naturalistic data set, considering syntactic, semantic and pragmatic aspects from both textual and visual language perspectives. 

The results draw attention to numerous design opportunities emphasising the multimodal resources that are relevant in this context. These considerations extend the potential for design solutions in an area that has recently placed more emphasis on speech interaction as the primary target for natural language interfaces. By considering this kind of everyday communication from the perspective of end-user programming, we can see a variety of ways in which speech interaction with the IoT might be extended, as already demonstrated in the familiar but surprisingly rich domain of the sticky note.

\vspace{-5pt}

\subsubsection{Acknowledgement}
{\small This work is supported by a Swiss National Science Foundation Early Postdoc Mobility fellowship (\#P2TIP2\_152264), and is also partially funded by International Alliance of Research Universities (IARU)Travel Grant and The ANU Vice Chancellor Travel Grant.}
\vspace{-5pt}

\bibliographystyle{splncs}
\bibliography{main}

\end{document}